\documentclass[graybox]{svmult}

\usepackage{type1cm}        
\usepackage{makeidx}         
\usepackage{graphicx}        
\usepackage{multicol}        
\usepackage[bottom]{footmisc}

\usepackage{newtxtext}       %
\usepackage[varvw]{newtxmath}       

\usepackage[tableposition=top,font=small,skip=5pt]{caption}

\usepackage{xcolor} 

\newcommand{\mypar}[1]{\noindent\textbf{#1}}

\usepackage{hyperref}
\usepackage{booktabs}
\usepackage{float}

\usepackage{algorithm}
\usepackage{algpseudocode}

\usepackage[normalem]{ulem}

\usepackage{tikz}
\usetikzlibrary{shapes.geometric, arrows}

\tikzstyle{startstop} = [rectangle, rounded corners, minimum width=2cm, minimum height=1cm,text centered, draw=none, fill=none]
\tikzstyle{process} = [rectangle, minimum width=3cm, minimum height=1cm, text centered, draw=black, fill=orange!30]
\tikzstyle{arrow} = [thick,->,>=stealth]

\makeindex     

\begin{document}

\title*{Effectiveness of Adversarial Benign and Malware Examples in Evasion and Poisoning Attacks}
\titlerunning{Comparison of Adversarial Benign and Malware Adversarial Examples}
\author{Matou\v{s} Koz\'{a}k\orcidID{0000-0001-8329-7572} and\\ Martin Jure\v{c}ek\orcidID{0000-0002-6546-8953}}
\institute{Matou\v{s} Koz\'{a}k \at Faculty of Information Technology, Czech Technical University in Prague, Prague, Czechia, \email{matous.kozak@fit.cvut.cz}
\and Martin Jure\v{c}ek \at Faculty of Information Technology, Czech Technical University in Prague, Prague, Czechia, \email{martin.jurecek@fit.cvut.cz}}

\maketitle


\abstract{Adversarial attacks present significant challenges for malware detection systems. This research investigates the effectiveness of benign and malicious adversarial examples (AEs) in evasion and poisoning attacks on the Portable Executable file domain. A novel focus of this study is on benign AEs, which, although not directly harmful, can increase false positives and undermine trust in antivirus solutions. We propose modifying existing adversarial malware generators to produce benign AEs and show they are as successful as malware AEs in evasion attacks. Furthermore, our data show that benign AEs have a more decisive influence in poisoning attacks than standard malware AEs, demonstrating their superior ability to decrease the model's performance. Our findings introduce new opportunities for adversaries and further increase the attack surface that needs to be protected by security researchers.}

\section{Introduction} \label{sec:introduction}
Malware, an abbreviation for malicious software, refers to a wide range of harmful software types, including viruses, worms, trojans, ransomware, and spyware. These harmful programs are intended to cause damage, disruption, or illegal access to computer systems, posing serious risks to individuals, companies, and national security \cite{gerencer2020}.

To tackle these threats, security engineers design malware detection systems, antiviruses (AVs), to detect and neutralize harmful behavior. Traditional malware detection solutions typically use signature-based methods that compare known patterns of harmful code to a database of signatures \cite{al2019review}. While these methods are successful against known threats, they fail to detect new, previously unknown malware patterns. To overcome this issue, heuristic and behavior-based detection algorithms have been developed. Heuristic-based methods evaluate software code structure and behavior to identify potentially dangerous actions \cite{bazrafshan2013survey}, whereas behavior-based methods monitor program runtime activity to find anomalies that indicate malicious actions \cite{nachenber1997emulator,jacob2008behavioral}. More recently, machine learning (ML) algorithms have improved malware detection by recognizing patterns and features in big datasets \cite{singh2021survey,comar2013combining,raff2017malware,anderson2018EMBER}.

As a result, attackers are constantly evolving their strategies to avoid discovery. For example, polymorphism \cite{drew2016polymorphic} or metamorphism \cite{szor2001hunting}, where malware encrypts or rewrites its code to appear differently to malware detectors. A new emerging threat in the malware domain is adversarial attacks \cite{podschwadt2019effectiveness}, which create inputs, so-called \textit{adversarial examples} (\textit{AEs}), that trick malware detectors into misidentifying malware as benign (non-malicious software also referred to as goodware).

In the field of malware detection, AEs are purpose-engineered inputs to incentivize the attacked model to make incorrect predictions. These perturbations are often subtle, yet they significantly impact the model's performance \cite{demetrio2019explaining}. In the malware area, adversarial attacks fall into two categories: evasion attacks and poisoning attacks. Evasion attacks entail constructing AEs that avoid detection by the model during inference \cite{anderson2018learning,kreuk2018deceiving,kozak2024creating,song2022mab,demetrio2021adversarial}. Poisoning attacks attempt to undermine the training process by contaminating the training dataset with AEs, to decrease the model's performance \cite{biggio2014poisoning,sasaki2019embedding}.

This work investigates the effectiveness of benign and malware AEs in evasion and poisoning attacks on Portable Executable (PE) files. A unique aspect of this work is the focus on benign AEs, which, unlike malicious counterparts, do not constitute a direct threat to the system. However, benign AEs can have a major impact on malware detection systems' performance by raising false positives. Consequently, this can lead to a loss of trust in AV products and unnecessary bottlenecks for legitimate software suppliers. To the best of our knowledge, this is the first work that compares the effectiveness of benign and malware AEs in evasion and poisoning attacks.

To summarize, our contributions are (i) we introduce and formally define the concept of benign AEs; (ii) we propose how contemporary generators of adversarial malware can be modified to create benign AEs; (iii) we show comparable effectiveness of malware and benign AEs in evasion attacks; (iv) we demonstrate superior influence of benign AEs in poisoning attacks over traditional malware AEs.


\section{Background} \label{sec:background}
In this section, we introduce the necessary background for this work. We start by explaining the concepts of adversarial machine learning while focusing on two main adversary scenarios in the malware domain: evasion and poisoning attacks. We follow with a description of the PE file format.

\subsection{Adversarial Machine Learning} \label{sec:adversarial_ML}
\textit{Adversarial machine learning} is a field that studies enhancing ML systems' resilience to adversarial attacks from both the outside (evasion attacks) and the inside (data poisoning). An adversarial attack is a well-planned action designed to deceive the ML model. The victim model is also known as a target model, whereas the attacker is also referred to as an adversary. An \textit{adversarial example} (\textit{AE}) is the object that is used to conduct the adversarial attack, e.g., a modified malware sample that evades detection or a tampered data point hidden in the training dataset. The next part outlines the taxonomy of adversarial attacks in the domain of malware detection and potential defense techniques.

We use the taxonomy offered by Huang et al. \cite{huang2011adversarial} as it is one of this topic's most comprehensive and security-related descriptions. Adversarial attacks are distinguished by three major characteristics: influence, security violation, and specificity. \\

\mypar{Influence.} The first attribute represents opponents' capacity to attack a particular model. The first type is termed \textit{causative} attacks, where the adversary may affect the training process of the model, e.g., conceal incorrectly labeled samples into the training dataset (data poisoning). The second type is \textit{exploratory} attacks. These attacks do not affect the training process, and their purpose is to learn about the model and avoid its detection measures, such as a modified malicious file that evades detection (evasion attack).

\mypar{Security violation.} The second attribute describes the type of security breach committed by the attacker. If an adversarial attack increases the model's false negative rate (adversarial malware samples classed as benign), we term it an \textit{integrity} attack. An \textit{availability} attack occurs when an attack increases both false negative and false positive rates, rendering the model unsuitable for any prediction. The last kind is a \textit{privacy} attack, also known as a model-stealing attack, which aims to steal the model's sensitive information, such as the training dataset or model parameters.

\mypar{Specificity} The third attribute represents the scope of the adversarial attack. Assume the attack is directed at a limited and specified subset of samples, we mark it as a \textit{targeted} attack. On the other hand, an \textit{indiscriminate} attack is a scenario where any sample can be misclassified. \\

We continue with a more detailed description of evasion and poisoning attacks, followed by common defense measures deployed against adversarial attacks.

\subsubsection{Evasion Attacks} \label{sec:background_AML_evasion_attacks}
An adversary may alter the input data in an evasion attempt to avoid being discovered by the detector. Evasion attacks, as they relate to malware detection, entail the development of malware that may evade the classifier and remain undetected.

Let $f: \mathbb{R}^n \rightarrow \{0, 1\}$ be a binary classifier where $f(x) = 1$ indicates a malicious sample and $f(x) = 0$ indicates a benign sample. An evasion attack seeks to find an adversarial malware example $x_{adv}$ such that

\begin{align*}
 x_{adv} = x + \delta
\end{align*}

\noindent where $\delta$ is a perturbation added to the original input $x$, and $f(x_{adv}) = 0$ while $f(x) = 1$. To prevent suspicion or behavioral changes, minimizing the perturbation $\delta$ to maintain $x_{adv} $'s similarity to $x$ is a common practice.

To formally define the concept of adversarial benign examples, we follow the same principles as with malware AEs but modify the added perturbation $\delta$ such that the $f(x_{adv}) = 1$ while $f(x) = 0$.

Numerous techniques are used for creating $x_{adv}$ examples. We follow with a brief introduction of a selected few and a more detailed case study in the later sections of this work. \\

\mypar{Feature Manipulation.} Attackers change particular aspects of the feature vector representing malware such that the classifier considers it safe software. This attack usually involves the attacker having good knowledge about what feature representation the target classifier uses to be able to successfully craft an evasive malware AE. The critical challenge for this attack is to devise an algorithm for transforming the adversarial feature vector back to executable binary format so that the AE can be deployed outside of the laboratory settings \cite{yilmaz2020improving,hu2017generating}.

\mypar{Adversarial EXEmples.} To overcome the problem of mapping from feature vectors to binary executables, attackers can create AEs by introducing noise or perturbations directly to binary code. This technique does not require knowledge about the inner workings of the target classifier as it can be used in black-box settings where only the prediction label is used as feedback for the adversary. However, creating both evasive and functional AEs is a complex problem where a good knowledge of the executable's binary format is needed \cite{kozak2024creating,demetrio2021functionality}.

\mypar{Obfuscation.} To keep the classifier from realizing the true nature of the presented software, methods like packing and code obfuscation are used. Obfuscation involves transforming the malware's code into a form that is difficult to analyze, whereas packing compresses and encrypts the code to prevent detection \cite{etter2023evading,anderson2018learning}.

\subsubsection{Poisoning Attacks}
The goal of poisoning attacks is to modify the training dataset that is utilized to train the detector. In terms of malware detection, this can consist of camouflaging AEs inside the training set to hinder the model's training process and consequently mistakenly identify some malware as benign or vice versa.

Let $D = {(x_i, y_i)}_{i=1}^M$ be the training dataset where $x_i$ represents the input features and $y_i$ represents the corresponding labels. A poisoning attack seeks to modify the training dataset by including poisoned samples such that classifier $f’$ trained on poisoned dataset $D’$ prediction behavior is changed

\begin{align*}
f’(x) \neq f(x)
\end{align*}

\noindent for some input $x$, causing the model to make incorrect predictions.



Among the frequent techniques used to create poisoned samples are data injection and label manipulation. \\

\mypar{Data Injection.} Attackers can influence the model's learning process by adding deliberately constructed harmful samples to the training data. These samples are meant to appear genuine, yet they have subtle details that cause the model to be misled. The samples can be crafted using a gradient ascent strategy to optimize the impact of the poisoned samples against a specific target detector \cite{biggio2012poisoning}.

\mypar{Label Manipulation}.To fool the model, existing samples in the training set can have their labels flipped. For instance, labeling malware samples as benign can cause the model to learn incorrect associations between features and labels. The adversary's goal is to determine which samples' labels to change to maximize the influence on the training stage \cite{xiao2012adversarial}.



\subsubsection{Defense Against Adversarial Attacks}
The goal of adversarial machine learning is to mitigate the risks posed by adversarial attacks. Researchers have created various protection methods to make ML models more resilient and improve reliability and confidence in their decisions. However, the trade-off between model robustness and performance must be carefully controlled to guarantee that the detection system remains efficient and accurate. \\

\mypar{Adversarial Training.} Adding correctly labeled AEs to the training set is known as adversarial training. The model gains the ability to identify and reject adversarial inputs by using these instances throughout the training phase. This technique can strengthen the model's resistance to evasive attacks \cite{madry2018towards,lucas2023adversarial}.

\mypar{Data Sanitization.} Methods for data sanitization, such as detecting anomalies, can be applied to detect inputs that substantially diverge from the trusted training set. Through the system's ability to identify questionable inputs, AEs can be excluded from the system, preventing both evasive and poisoning attacks \cite{quiring2020against, abusnaina2023burning}. An example of data sanitization is the $L2$ defense (also called sphere defense) technique where data points are projected onto a high dimensionality sphere, and points beyond the sphere's radius are excluded \cite{koh2022stronger}.

\mypar{Feature Representation.} The complexity and attack surface disposable for the attacker can be reduced by increasing the robustness of the feature representation used by the model. For example, decreasing the precision of individual features \cite{xu2017feature} or dimensionality reduction \cite{bhagoji2018enhancing} can flaw attackers' chances of bypassing the detection. Additionally, domain knowledge in devising the feature representation is critical as including unrelated features can mislead the model in learning false connections that the adversaries can exploit \cite{demetrio2019explaining}.

\mypar{Robust Model Architecture.} The security of malware detection systems can be increased by creating model designs that are inherently resistant to adversarial attacks. For example, using multiple classifiers \cite{rashid2023effectiveness} or plug-in adversary detectors \cite{kozak2024updating} can increase the difficulty of executing a successful adversarial attack.

\subsection{Portable Executable File Format} \label{sec:pe_file_format}
The \textit{Portable Executable} (\textit{PE}) file format is a data format that stores the information required by the Windows operating system loader to manage the executable code. It is used to store executable (EXE), object code, dynamic link libraries (DLL), and other files on both 32-bit and 64-bit Windows operating systems \cite{microsoftPE}.

The structure of the PE file format can differ slightly depending on which type of file it represents. This section focuses on the PE file format structure used for EXE files. The format is organized as follows: \\

\mypar{MS-DOS Header.} Every PE file begins with the MS-DOS header, which is a 64-byte structure that converts the PE file into MS-DOS executable. This header contains a magic number that indicates the file is MS-DOS executable. At the end of the header is an offset of the COFF file header.

\mypar{MS-DOS Stub.} The MS-DOS header is followed by the MS-DOS stub, a short MS-DOS program that typically prints a message such as ``This program cannot be run in DOS mode'' if the executable is run on MS-DOS.

\mypar{COFF File Header.} Next, the COFF File header is located at the offset found in the MS-DOS header. Before the actual COFF header appears, a 4-byte signature field identifies the file as a PE file with a value of \verb|PE\0\0|. The following 20 bytes contain generic information about the PE file, e.g., machine type, timestamp, or number of sections.

\mypar{Optional Header.} Following is the Optional header. For EXE files, the header includes essential information for the OS loader, such as the entry point address, linker version, image base, and section alignment.

\mypar{Section Headers.} The Section headers come right after the optional header, with each header totaling 40 bytes of section description: name, virtual size and address, section attributes, and more.

\mypar{Section Data.} Following the table of section headers is the actual section content, including code and other resources. Typical sections and their content are .text (executable code), .data (initialized data), .rdata (read-only data), .debug (debugging information), and .idata (imported libraries and functions).

\section{Generators of Adversarially Modified Software} \label{sec:adversarial_generators}
In this section, we dive into what the generators of adversarial malware are and portray how some contemporary generators work. Finally, we propose the notion of adversarial benign generators and how we can modify contemporary generators of adversarial malware to create benign AEs.


\subsection{Generators of Adversarial Malware} \label{sec:adversarial_malware_generators}
The purpose of adversarial malware generators is to produce malware samples capable of evading detection by security systems, especially those that employ ML-based models. Adversarial malware generators are primarily used to test and enhance malware detection systems' resilience. Researchers and security experts can find flaws in their detection methods and create more robust defenses by testing the systems on adversarial examples. However, these generators can also be abused by bad actors to produce malware that hides from detection, which poses potential security issues.

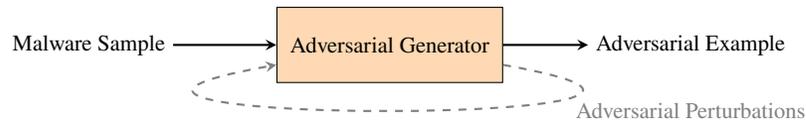
\begin{figure}[!htb]
    \centering
\begin{tikzpicture}[node distance=4cm]

    \node (input) [startstop] {Malware Sample};
    \node (generator) [process, right of=input] {Adversarial Generator};
    \node (output) [startstop, right of=generator] {Adversarial Example};
    
    \draw [arrow] (input) -- (generator);
    \draw [arrow] (generator) -- (output);

    \draw [arrow, dashed, gray] (generator) edge [out=350,in=-170,loop,looseness=4] node[right of=edge] {\footnotesize{Adversarial Perturbations}} (generator);
\end{tikzpicture}
    \caption{Workflow of adversarial malware generators.}
    \label{fig:adversarial_malware_generator}
\end{figure}

From the black-box point of view as portrayed in Figure \ref{fig:adversarial_malware_generator}, the generators work by taking a genuine malware sample as an input, followed by an application of adversarial perturbations and finally producing a modified, so-called adversarial example.

The following is a selection of adversarial malware generators used in this work. We decided to select generators that do not require previous training. As such, we can use the same generator to create and fairly compare adversarial malware and benign examples. \\

\mypar{AMG.} The Adversarial Malware Generator (AMG) is a generator utilizing a reinforcement learning (RL) algorithm called proximal policy optimization (PPO) agent trained to apply a set of functionality-preserving modifications to previously detected samples. The modifications were carefully designed and thoroughly tested to maximize the functionality preservation of used input samples. The same modifications can also be used in random settings (without previous training), and the resulting AEs are highly evasive against commercial AVs \cite{kozak2024creating}. 

\mypar{FGSM.} In contrast to the original attack \cite{goodfellow2015explaining} utilizing the fast gradient sign method (FGSM) for the image domain, only short sequences of bytes (payloads) are adversarially perturbed. At first, random bytes are placed into unused space between sections or at the end of the file to ensure that the original functionality remains intact. The FGSM technique is then used to perturb only these sequences, misleading the target classifier \cite{kreuk2018deceiving}.

\mypar{GAMMA.} The Genetic Adversarial Machine Learning Malware Attack (GAMMA) is a generator that injects benign content at the end of the file or into newly created sections. The injected benign content is optimized using a genetic algorithm constrained to maximize evasion rate while minimizing the magnitude of the perturbation \cite{demetrio2021functionality}.

\mypar{MAB-Malware.} A RL-based generator using a multi-armed bandit (MAB) agent together with a set of macro and micro manipulations devised to maximize evasion with minimal perturbation. The generator works in two phases. First, the MAB agent applies a sequence of modifications until evasion is achieved. Subsequently, each applied modification is tested to be expendable and removed if found so. This minimization process is possible because the MAB agent does not imply any connection between applied actions, hence making it possible to remove some of them afterward \cite{song2022mab}.

\mypar{Partial-, Full-, Extend-DOS Manipulators.} Set of MS-DOS manipulators utilizing gradient-based optimization to adversarially modify malware's MS-DOS header and stub program. The \textit{Partial-DOS} generator modifies only the content of the MS-DOS header between the magic number and offset of the COFF File header \cite{demetrio2019explaining}. The \textit{Full-DOS} generator extends the modifications of Partial-DOS by perturbing the MS-DOS stub program as well \cite{demetrio2021adversarial}. Lastly, the \textit{Extend-DOS} generator further extends the modification beyond the end of the MS-DOS stub program until the beginning of the COFF File header \cite{demetrio2021adversarial}.

\subsection{Generators of Adversarial Goodware} \label{sec:adversarial_benign_generators}
While contemporary research in the adversarial machine learning and malware detection domain focuses only on the efficacy of malware AEs, we propose a novel approach that involves the creation of benign AEs. This new type of AE represents harmless software files deliberately compromised to be misclassified as malware. While not directly harmful, this strategy can dramatically influence the functioning of malware detection systems by increasing false positives, leading to a loss of trust in antivirus solutions and causing legitimate software vendors to face costly blockages. The following is an approach we use to modify the above-mentioned generators of adversarial malware to create effective benign AEs.

A common theme across generators of malware AEs is a stopping condition that stops the generating process when the target classifier $f$ no longer detects the malware sample. First, we must change this condition to a reverse scenario where the process is halted when AE is no longer classified as benign. This change is demonstrated in Algorithm \ref{algo:stopping_condition_benign_AEs}.

\begin{algorithm}[!htb]
    \caption{Stopping condition for generators of AEs.}
    \begin{algorithmic}[1]
        \Require $x_{\text{orig}}$: Original sample, $f$: Target classifier, $N_{\text{max}}$: Maximum number of iterations
        \Ensure $x_{\text{AE}}$: Adversarial example
        \State $x_{\text{AE}} \gets x_{\text{orig}}$
        \For{$i = 1$ to $N_{\text{max}}$}
            \If{$f(x_{\text{AE}}) \neq$ \sout{$malicious$} $benign$}
                \State \textbf{break}
            \EndIf
            \State Adversarially perturbe $x_{\text{AE}}$
        \EndFor
        \State \Return $x_{\text{AE}}$
    \end{algorithmic}
    \label{algo:stopping_condition_benign_AEs}

\end{algorithm}

Further, for RL-based generators AMG and MAB-Malware, we must prepare malware-looking content used to inject into unused spaces of PE files. We extract malware section content using the \verb|process_benign_dataset.py|\footnote{\url{www.github.com/weisong-ucr/MAB-malware/blob/main/process_benign_dataset.py}} script from the authors of MAB-Malware \cite{song2022mab} on our dataset of malware executables described later in Section \ref{sec:evaluation_setup}.

For gradient-based generators FGSM, PartialDOS, FullDOS, and ExtendDOS, the adversarial modifications need to be designed in such a manner that the perturbation calculated using the gradient of the target classifier's loss function moves the sample closer to the malware class. A simple solution can be achieved by changing the sign of the computed gradient, which is used to optimize the perturbation, hence reversing the direction of the movement introduced by the generator.

For the GAMMA generator, we must provide malware EXEs instead of benign files for the generator. Additionally, the fitness value function must be changed to prefer individuals that maximize the target classifier's prediction score, i.e., have a higher malicious confidence score.

The proposed changes are easy to implement and significantly increase the capabilities of existing generators (see \nameref{sec:evaluation}). We implement these changes in the source codes provided by the authors of AMG\footnote{\url{www.github.com/matouskozak/AMG}}, MAB-Malware\footnote{\url{www.github.com/weisong-ucr/MAB-malware}} and SecML Malware\footnote{\url{www.github.com/pralab/secml_malware}} (FGSM, GAMMA, PartialDOS, FullDOS, ExtendDOS generators).
    



    







\section{Experimental Evaluation} \label{sec:evaluation}
In this section, we present our experimental evaluation of the effectiveness of adversarial benign and malware examples. We start by describing our hardware setup, dataset, and used evaluation metrics. We follow by presenting our experimental comparison of benign and malware AEs in evasion and poisoning attacks. We conclude this section by discussing the results and pointing out the limitations of our research.

    
    

\subsection{Setup} \label{sec:evaluation_setup}
\mypar{Hardware.} Experiments presented in this work were conducted on the NVIDIA DGX Station A100 server. The server contains a single AMD 7742 CPU with 64 cores, four NVIDIA A100 GPUs, and 512 GB of system memory. However, both the generation of adversarial examples and subsequent experimentation can be reproduced on a standard personal computer with at least 16 GB of system memory.

\mypar{Dataset.} We use three datasets for the experiments. Benign EXE files were obtained from a clean Windows 10 installation. Next, we downloaded malicious EXE files from the VirusShare\footnote{\url{www.virusshare.com}} data repository, whom we thank for access. We verified malware and benign EXEs to be truly malicious and harmless, respectively, by using the VirusTotal\footnote{\url{www.virustotal.com}} API and discarding samples classified as the contrary class. Lastly, we also use the EMBER dataset \cite{anderson2018EMBER} to extend our range of available samples. Namely, we use parts of the EMBER training dataset to balance the ratio of benign to malware files in the training sets used in the experimental evaluation. Additionally, we benefit from the recognized EMBER test dataset and use it as our default evaluation set for poisoning attacks. The EMBER test set contains 200000 samples, equally distributed between malicious and benign classes.

\subsection{Evaluation Metrics}
To assess the effectiveness of evasion and poisoning adversarial attacks, several key metrics are used. These metrics contribute to quantifying the success of adversarial attacks and the resilience of protection measures. The following are the metrics used: \\

\mypar{Confusion Matrix.} The base of statistical evaluation is the confusion matrix that thoroughly describes the performance of the studied model. The table is structured as follows:

\begin{table*}[!htb]
    \centering
    \begin{tabular}{l|l|l}
        \hline
        \textbf{}                & \textbf{Predicted Positive} & \textbf{Predicted Negative} \\ \hline
        \textbf{Actual Positive} & True Positive (TP)          & False Negative (FN)         \\ \hline
        \textbf{Actual Negative} & False Positive (FP)         & True Negative (TN)          \\ \hline    
    \end{tabular}

\end{table*}

\begin{itemize}
    \item \textbf{True Positive (TP)} The number of correctly detected malicious samples.
    \item \textbf{True Negative (TN)} The number of correctly identified harmless files.
    \item \textbf{False Negative (FN)} The number of undetected malware files.
    \item \textbf{False Positive (FP)} The number of incorrectly blocked benign files.
\end{itemize}

\mypar{Detection Rate (DR).} The proportion of correctly detected malicious samples to the total number of malware samples, commonly referred to as sensitivity or true positive rate (TPR).

\begin{equation*}
    DR = \frac{TP}{TP + FN}
\end{equation*}

\mypar{False Positive Rate (FPR).} The proportion of benign files classified as malicious to the total number of benign samples.

\begin{equation*}
    FPR = \frac{FP}{FP + TN}
\end{equation*}

\mypar{Evasion Rate (ER).} The proportion of adversarial files that bypassed the detector to the total number of adversarial samples. Note that this metric can be used both for benign and malware AEs.

\begin{equation*}
    ER = \frac{\text{number of missed AEs}}{\text{total number of AEs}}
\end{equation*}

\mypar{Contamination Rate (CR).} A metric used in the context of poisoning attacks representing the proportion of training dataset that has been compromised (poisoned).

\begin{equation*}
    CR = \frac{\text{number of poisoned samples}}{\text{total number of training samples}}
\end{equation*}

\mypar{Receiver Operating Characteristics (ROC) Curve} A graphical representation of classifier's performance across different threshold settings. The thresholds are displayed based on the values of FPR against TPR.



\subsection{Evasion Attack Against the Target Classifiers} \label{sec:experiment_evasion_attack}
In the first experiment, we generate adversarial malware examples using the generators described in Section \ref{sec:adversarial_malware_generators} and adversarial benign examples by their modified versions described in Section \ref{sec:adversarial_benign_generators}. We use the previously described datasets of malware and benign executables as inputs for the generators. Next, we compare the genuine samples with the adversarial counterparts and discard examples that do not contain adversarial perturbations. Several factors can cause the lack of modification, the most common being that the genuine sample was already misclassified (either as benign for malware AEs or as malware for benign AEs) by the target classifier before the modification process. As a target classifier, we use the default selection done by the authors of the generators: gradient boosted decision tree (GBDT) \cite{anderson2018EMBER} and MalConv, a convolutional neural network classifier \cite{raff2017malware}. The resulting counts of used benign and malware AEs are shown in Table \ref{table:dataset_stats}, where each row represents a single generator with the target classifier specified in parenthesis after the generator name.

\begin{table}[!htb]
    \centering
    \caption{Sums of generated benign and malware AEs for each generator after filtering.}
    \begin{tabular}{@{}l|cc@{}}
    \toprule
    Generator             & Benign & Malware \\ \midrule
    AMG-random (GBDT)     & 3158   & 6595    \\
    ExtendDOS (MalConv)   & 1566   & 5511    \\
    FGSM (MalConv)        & 1321   & 5090    \\
    FullDOS (MalConv)     & 1568   & 2035    \\
    GAMMA (MalConv)       & 3132   & 5506    \\
    MAB-Malware (GBDT)    & 1439   & 6614    \\
    MAB-Malware (MalConv) & 1477   & 5397    \\
    PartialDOS (MalConv)  & 1568   & 3065    \\ \midrule
                          & \textbf{15229}  & \textbf{39813} \\ \bottomrule
    \end{tabular}
    \label{table:dataset_stats}

\end{table}

To evaluate the effectiveness of generated benign and malware AEs in evasion attacks, we compare the evasion rates against the respective target classifiers. Based on the taxonomy introduced in Section \ref{sec:adversarial_ML}, this represents a targeted exploratory attack. The results are shown in Table \ref{table:evassivness_AEs}, where each row represents a single generator, and benign and malware columns represent the evasion rates (in \%) of generated AEs against the target classifier. Based on the results, we do not have a clear pattern of whether benign or malware AEs are more effective against MalConv or GBDT detectors. The ExtendDOS, FullDOS (MalConv), and MAB-Malware (GBDT) generate highly evasive benign AEs with evasion rates between 84\%-97\%. The rest of the generators are more successful in creating evasive malware AEs with MAB-Malware and PartialDOS (MalConv), generating between 90\%-93\% of evasive samples.

\begin{table}[!htb]
    \centering
    \caption{Evasion rates of malware and benign AEs from different generators against the target classifier for which the samples were generated. [\%]}
    \begin{tabular}{@{}l|cc@{}}
    \toprule
                         & benign           & malware \\ \midrule
    AMG-random (GBDT)    & 30.34            & \textbf{36.12} \\
    ExtendDOS (MalConv)  & \textbf{96.81}   & 43.51       \\
    FGSM (MalConv)       & 30.96            & \textbf{42.16}       \\
    FullDOS (MalConv)    & \textbf{97.64}   & 56.02       \\
    GAMMA (MalConv)      & 16.32            & \textbf{36.29}       \\
    MAB-Malware (GBDT)   & \textbf{84.09}   & 74.77       \\
    MAB-Malware (MalConv)& 65.0             & \textbf{90.99}       \\
    PartialDOS (MalConv) & 71.88            & \textbf{93.05}       \\ \bottomrule
    \end{tabular}
    \label{table:evassivness_AEs}

\end{table}

\subsection{Poisoning Attacks Against the GBDT Classifier}
We situate the poisoning attack as an indiscriminate causative attack, i.e., our goal is to mislead the subsequently trained model into any misclassification. Additionally, we do not generate poisoning samples separately but use the previously crafted adversarial examples from Section \ref{sec:experiment_evasion_attack}. Note that we use the EMBER train samples to represent the non-poisoned samples in the training dataset over the genuine executables used to generate the poisoning samples. This represents a scenario where the attacker does not possess knowledge of the samples that are already present in the training dataset. As a victim model, we choose the GBDT model trained using LightGBM library \cite{lightgbm, ke2017lightgbm}. The input binary files are represented by a 2381-long feature vectors containing information extracted from the PE files \cite{anderson2018EMBER}. We evaluate the trained GBDT classifier on the EMBER test set.

\subsubsection{Poisoning by Single AE Generator} \label{sec:experiment_poisoning_single}
In the second experiment, we investigate the effectiveness of individual generators in the poisoned training dataset. We explore different ratios of dataset contamination ranging from 0\% (only genuine EMBER train samples) to 100\% (poisoned samples replace a single class of genuine samples).
During all experiments, we maintain an even balance between genuine and malware classes. Due to the limited number of generated benign AEs (see Table \ref{table:dataset_stats}) and to ensure a fair comparison between malware and benign AEs, we limit the maximum number of used AEs to 1000, capping the total size of the training dataset to 2000 samples.

\begin{figure}[!htb]
    \centering
    \includegraphics[width=\textwidth]{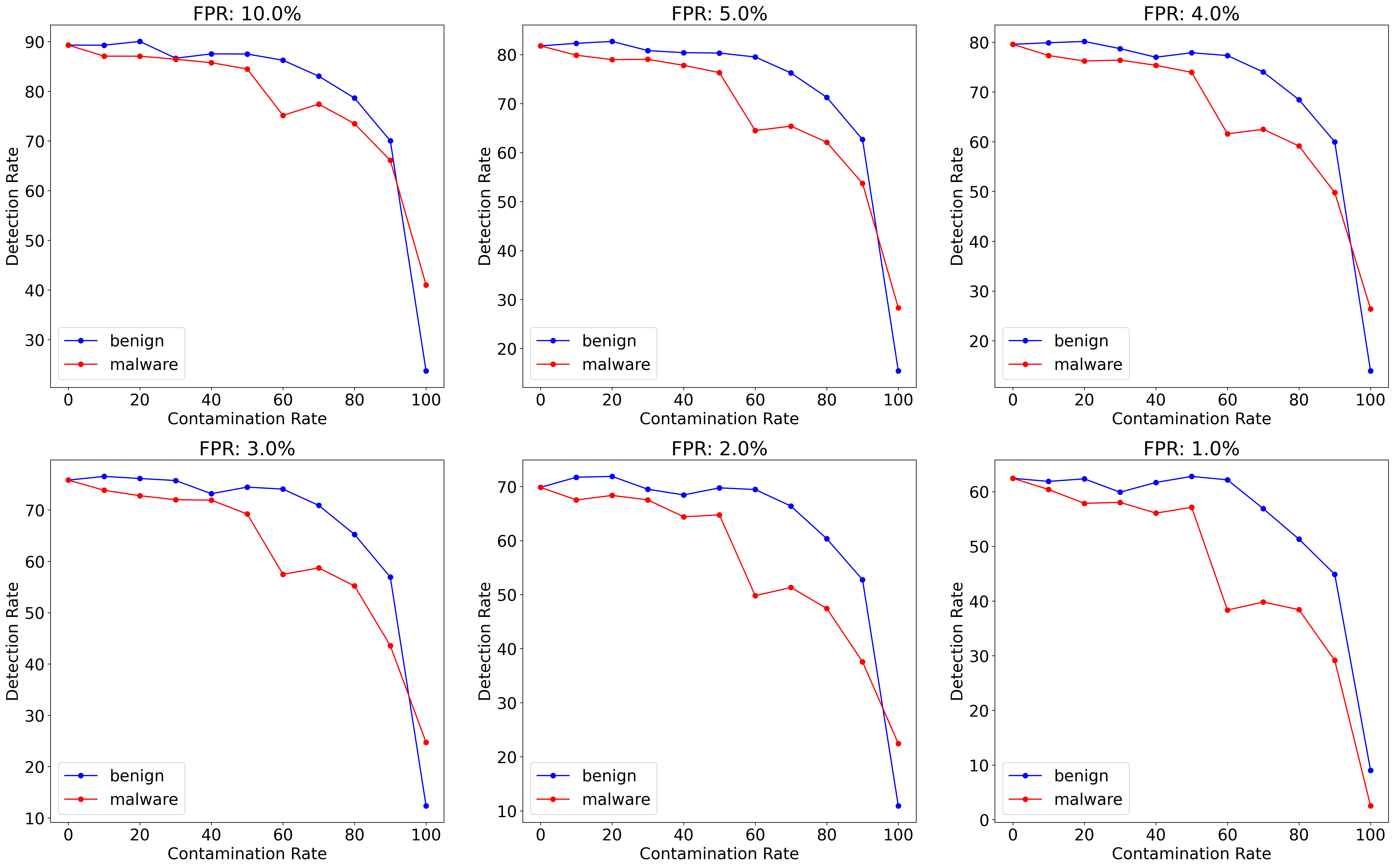}
    \caption{Comparison of detection rates at fixed levels of FPR after poisoning the dataset by malware or benign AEs from the MAB-Malware (GBDT) generator.}
    \label{fig:MAB_GBDT_AEs_separate_detection_rates_compare_ratios}
\end{figure}

Based on the results we collected, the effect of poisoning by benign or malware AEs varies significantly based on the generator used. For example, in Figure \ref{fig:MAB_GBDT_AEs_separate_detection_rates_compare_ratios}, where each subfigure represents a comparison between benign and malware AEs for a fixed level of FPR, we can see that malware AEs generated by MAB-Malware (GBDT) are more successful in poisoning the classifier's training dataset than the benign counterparts. The graphs show that for all evaluated levels of FPR and contamination rates, the presence of malware AEs in the training dataset detriments trained GBDT's detection rate more than benign AEs.

\begin{figure}[!htb]
    \centering
    \includegraphics[width=\textwidth]{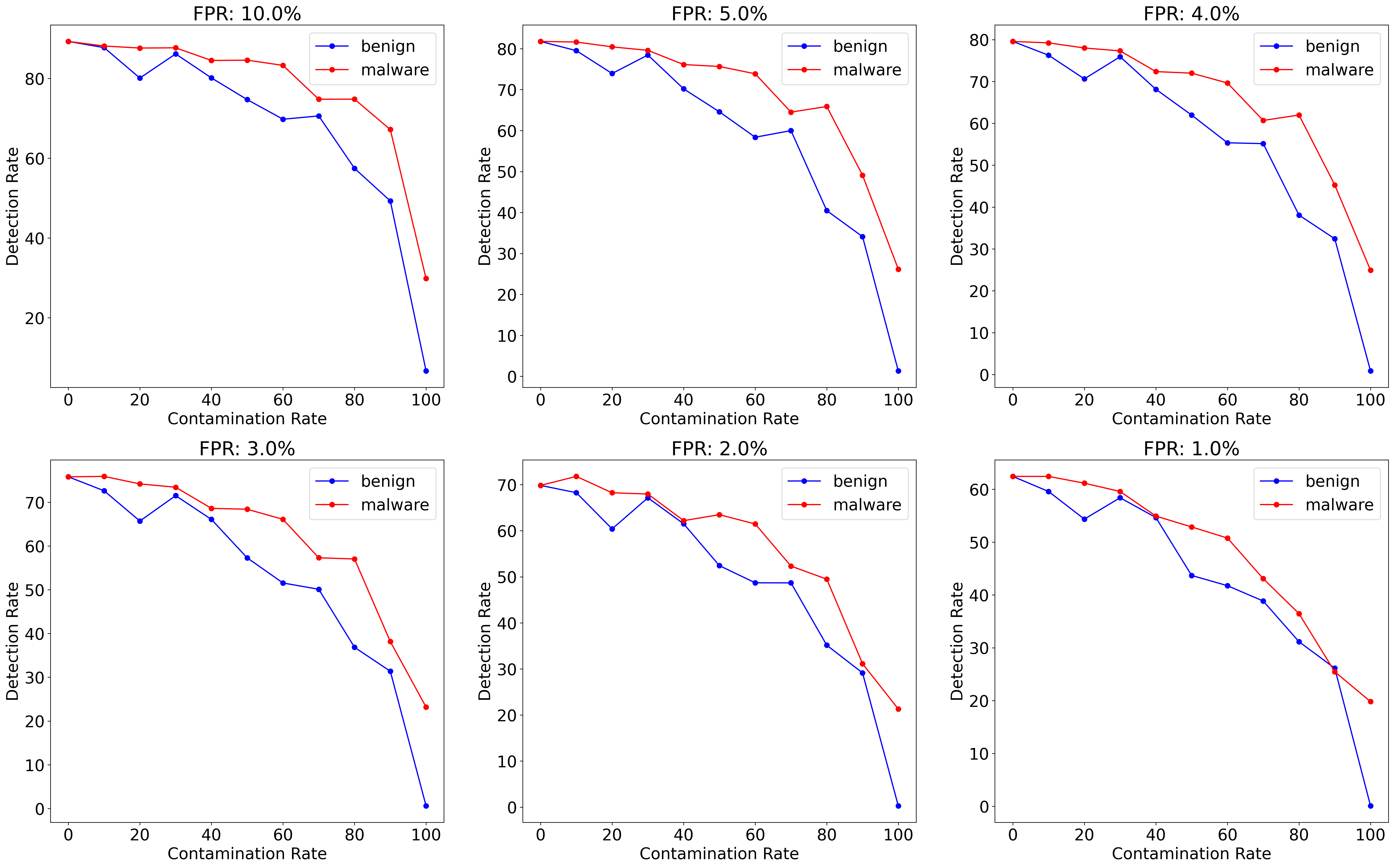}
    \caption{Comparison of detection rates at fixed levels of FPR after poisoning the dataset by malware or benign AEs from the FullDOS (MalConv) generator.}
    \label{fig:FullDOS_AEs_separate_detection_rates_compare_ratios}
\end{figure}

On the other hand, Figure \ref{fig:FullDOS_AEs_separate_detection_rates_compare_ratios} shows that the benign AEs created by FullDOS (MalConv) are significantly more potent in decreasing the detection rate of the GBDT model than malware AEs. 

\begin{figure}[!htb]
    \centering
    \includegraphics[width=\textwidth]{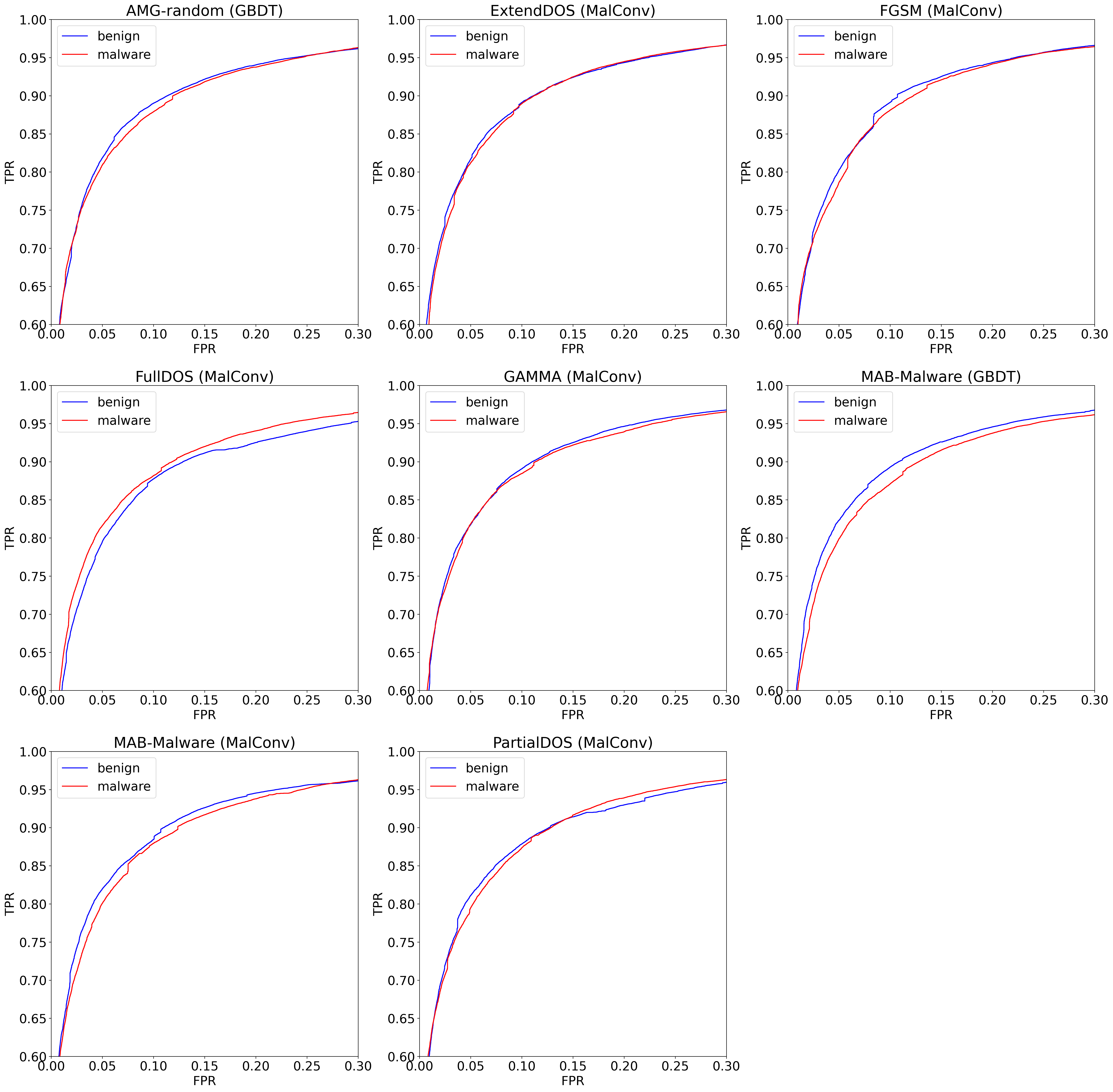}
    \caption{Comparison of ROC curves after poisoning 10\% of training dataset by malware or benign AEs from different generators.}
    \label{fig:separate_generators_roc_curves_0.1}
\end{figure}

The high variance in the success of benign and malware AEs can also be seen in Figure \ref{fig:separate_generators_roc_curves_0.1}, where we display ROC curves for each of the generators at a fixed contamination rate of 10\%. Notably, we can see that the effect of benign and malware AEs is similar, with only marginal differences across all evaluated generators. 


\begin{table}[!htb]
    \centering
    \caption{Comparison of detection rates at 1\% FPR for individual generators after poisoning 10\% of the training dataset. [\%]} 
    \begin{tabular}{@{}l|cc@{}}
    \toprule
    Generator             & Benign      & Malware \\ \midrule
    AMG-random (GBDT)     & 62.46       & \textbf{61.66}    \\
    ExtendDOS (MalConv)   & 63.7        & \textbf{61.64}    \\
    FGSM (MalConv)        & 60.49       & \textbf{59.84}    \\
    FullDOS (MalConv)     & \textbf{59.58}      & 62.44     \\
    GAMMA (MalConv)       & \textbf{60.92}       & 63.84    \\
    MAB-Malware (GBDT)    & 61.87       & \textbf{60.39}    \\
    MAB-Malware (MalConv) & 62.93       & \textbf{61.5}     \\
    PartialDOS (MalConv)  & \textbf{60.64}       & 61.4     \\ \bottomrule
    \end{tabular}
    \label{table:poisoning_contamination_rate-0.1_fpr-0.01}

\end{table}

A similar conclusion can be made by looking at Table \ref{table:poisoning_contamination_rate-0.1_fpr-0.01}, which presents the detection rates at fixed 1\% FPR after poisoning 10\% of the training dataset by benign or malware AEs from respective generators. We can see that PartialDOS, FullDOS, and GAMMA benign AEs lead to 59\%-61\% detection rates at 1\% FPR while the malware counterparts report higher detection rates of 61\%-64\% (a lower DR means more successful poisoning attack). The malware AEs from the rest of the generators are more effective (59\%-62\% DRs) than benign AEs (60\%-64\% DRs).

\subsubsection{Poisoning by Combination of AEs from Different Generators}
In the third experiment, we explore different scenarios based on the samples included in the training dataset. The first is a malware scenario where all generated malware AEs are accompanied by malware and benign samples from the EMBER dataset. The second is a benign scenario where all benign AEs are combined and extended with genuine samples from EMBER. The third is a mixture scenario, where both benign and malware AEs are combined together with genuine EMBER training samples. The total size of the final dataset for each scenario is limited to 30000 samples, with a balanced distribution between malicious and benign samples. As in the previous experiment, the range of adversarial contamination of the training dataset ranges from 0\% to 100\%, i.e., up to 15000 AEs are present in the training loop. 

\begin{figure}[!htb]
    \centering
    \includegraphics[width=\textwidth]{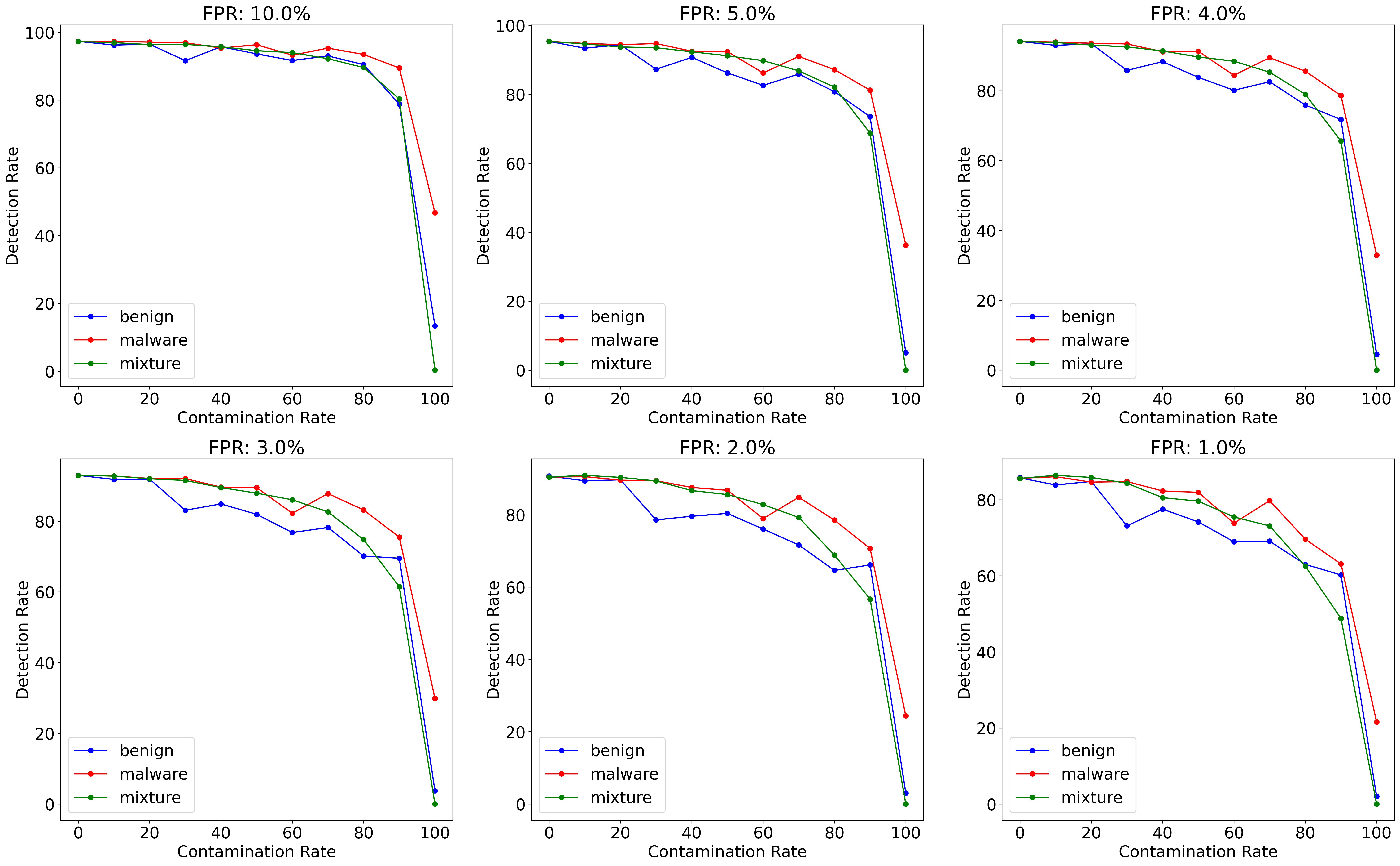}
    \caption{Comparison of detection rates at fixed FPR levels in malware, benign, and mixture scenarios.}
    \label{fig:combined_generators_detection_rates}
\end{figure}

The results are presented in Figure \ref{fig:combined_generators_detection_rates},  where each subfigure compares benign and malware AEs at a fixed level of FPR. Based on an initial look, benign AEs are causing more harm to the model's detection rate over malware or mixture scenarios with increasing contamination rate.

\begin{figure}[!htb]
    \centering
    \includegraphics[width=0.7\textwidth]{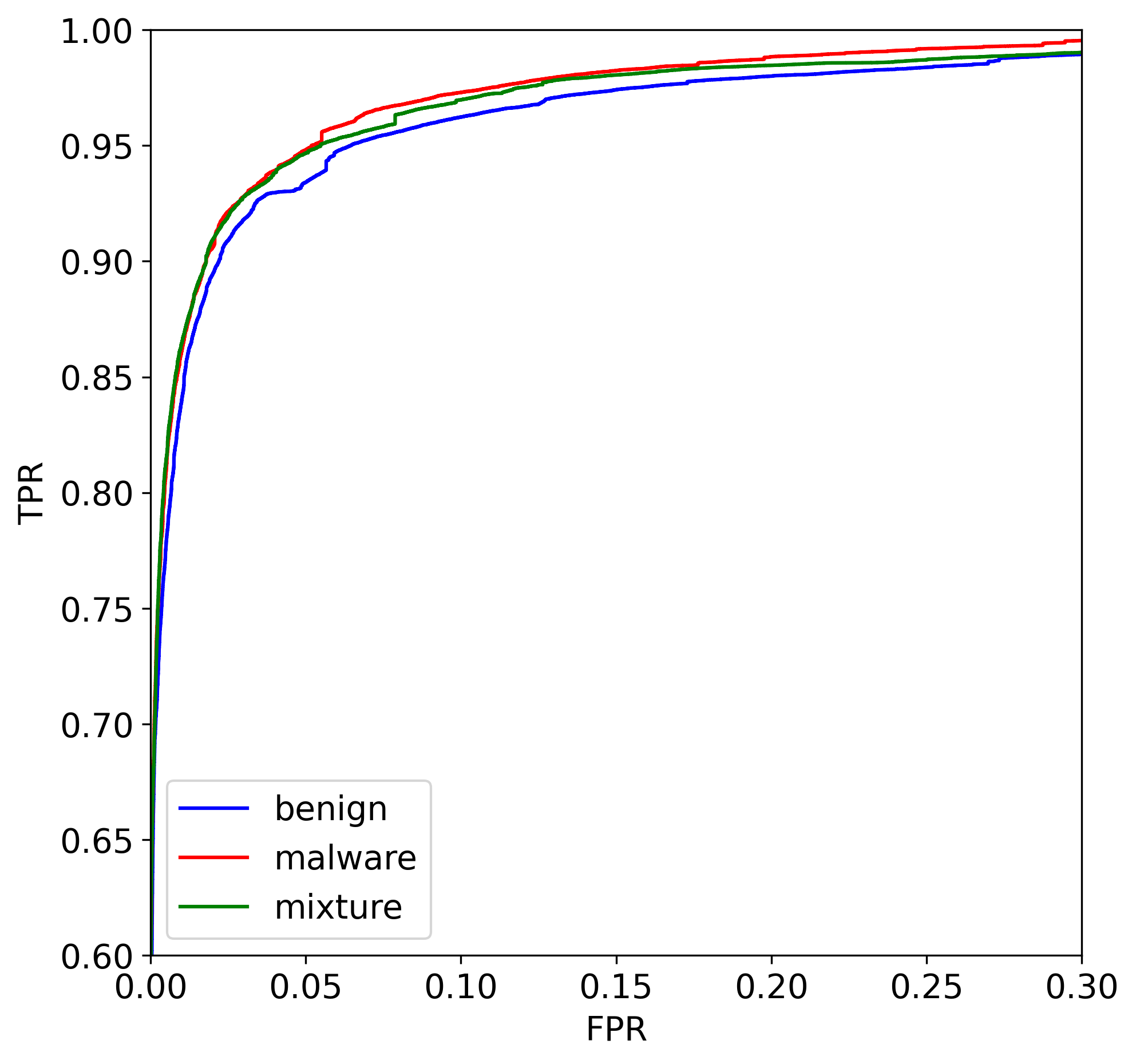}
    \caption{ROC curve after poisoning 10\% of training dataset by malware, benign or mixed AEs.}
    \label{fig:combined_generators_roc_curves_0.1}
\end{figure}

In more detail, we can see the performance at 10\% contamination rate in Figure \ref{fig:combined_generators_roc_curves_0.1}. From the ROC curve, we can see that poisoning the training dataset benign AEs from different generators consistently leads to the worst performance of the resulting trained model. The effect is apparent even for as little as 10\% of benign AEs in the training dataset for low levels of FPR (1\%-5\%). In Table \ref{table:poisoning_combined_CR-0.1}, we can see that for all tested levels of FPR, the presence of benign AEs in the training dataset decreases the detection rate by 0.96\%-2.19\% more than the corresponding scenario with malware AEs.

\begin{table}[!htb]
    \centering
    \caption{Comparison of detection rates at fixed levels of FPR in malware, benign, and mixture scenarios. [\%]} 
    \begin{tabular}{@{}l|ccc@{}}
        \toprule
        FPR & Benign & Malware & Mixture \\ \midrule
        1.0 & \textbf{83.88}   & 86.07        & 86.43        \\
        2.0 & \textbf{89.42}   & 90.58        & 90.92        \\
        3.0 & \textbf{91.8}    & 92.79        & 92.77        \\
        4.0 & \textbf{92.96}   & 93.92        & 93.81        \\
        5.0 & \textbf{93.41}   & 94.81        & 94.67        \\ \bottomrule
    \end{tabular}
    \label{table:poisoning_combined_CR-0.1}

\end{table}



\subsection{Discussion} \label{sec:experiments_discussion}
In evasion attacks, some generators are highly effective in crafting benign AEs that are successfully mispredicted as malware by the target classifier for which they were generated. Although, due to the nature of benign AEs, evasive benign AEs cannot cause harm to the attacked system, they can increase spikes in false positive reporting, consequently leading users to lose trust in the AV systems and vendors to complain due to blockage of their software.

In untargeted poisoning attacks, we do not see a significant difference in the effectiveness of contaminating the training dataset by benign or malware AEs crafted by a single generator. However, we still observe that benign AEs can be as effective as malware counterparts in poisoning attacks.

Moreover, using a combination of AEs from different sources for poisoning attacks, we report a measurable difference between benign and malware AEs. Notably, benign AEs are more effective in decreasing the detection rate at fixed FPR when included in the training dataset over malware (and mixed) AEs. This effectiveness presents a new opportunity for poisoning attacks where security researchers must keep focus on both malware and benign poisoning samples infiltrating the training datasets.

\mypar{Limitations.} One of the shortcomings of this paper is the size of the benign EXE dataset, which posed a limitation on the quality of the model used in the \nameref{sec:experiment_poisoning_single} experiment. This limitation caused the trained model to significantly underperform on the EMBER test set (62.44\% DR at 1\% FPR before poisoning attack), making it harder to compare the detrimental effect of poisoning by benign and malware AEs from a single generator. Another limitation of our work is the use of a sole GBDT classifier as a victim model for poisoning attacks. While several generators specifically target the GBDT classifier, many target the MalConv detector, which has a different architecture and is thus susceptible to different adversarial perturbations. Consequently, AEs against MalConv may not significantly affect the feature representation used for training GBDT, diminishing the effectiveness of created AEs.

\section{Related Work} \label{sec:related_work}
In this section, we present related research in adversarial machine learning and malware detection. We divide this section into two parts. The first part is focused on evasion attacks where adversaries generated AEs with the goal of bypassing detection by antivirus protection systems. The second part is dedicated to poisoning attacks where attackers infiltrate the training sets of malware detectors.

\subsection{Evasion Attacks}
Evasion attacks can be divided into groups based on their technique to generate AEs. We use the division into the three most prevalent groups in the domain of adversarial malware: gradient-based, reinforcement learning-based, and others.

The \textit{gradient-based} attacks take advantage of the back-propagation algorithm, commonly used for training deep neural networks \cite{goodfellow2015explaining,papernot2016limitations}. The techniques are based on injecting specially crafted perturbations that cause the target model to move its prediction in the direction of the gradient, thus decreasing its confidence in malware prediction.

Kolosnjaji et al. \cite{kolosnjaji2018adversarial} used gradient computation to adversarially perturb the overlay data of PE files, achieving a 60\% evasion rate against the MalConv detector.

Next, Kreuk et al. \cite{kreuk2018deceiving} presented an attack that injects up to 1000 bytes of adversarial content into unused regions of the PE file. Their attack misled the MalConv classifier in 99\% of cases, highlighting the severe vulnerability of pure ML-based antivirus systems to adversarial attacks.

Another attack on the MalConv detector was proposed by Demetrio et al. \cite{demetrio2019explaining}. The authors investigated which parts of the executable binary the MalConv model focuses on when making a prediction. Based on their results, the detector learned to use parts of the MS-DOS header to make its prediction decisions even though the MS-DOS header is currently included just for compatibility with older operating systems. Demetrio et al. exploited this finding and introduced an attack perturbing only the MS-DOS header and achieving an evasion rate of over 86\% against the MalConv.

The \textit{reinforcement learning-based} attacks use agents equipped with a set of manipulation actions on binary files. These agents are trained to apply these actions by continuously probing the target classifier and learning its inner decision-making \cite{anderson2018learning}.

The use of reinforcement learning agents for adversarial malware generation was pioneered by Anderson et al. \cite{anderson2018learning}. The authors deployed an actor-critic model trained to modify PE files. The agent was equipped with modifications such as adding new sections, packing, or including new imports. Their agent was able to bypass the GBDT detector in 24\% of cases.

Next, Song et al. \cite{song2022mab} used a state-less multi-armed bandit (MAB) agent to attack MalConv, GBDT, and commercial AVs. The MAB agent was armed with modifications such as adding new sections, appending benign content to overlay, or renaming current sections. The authors demonstrated an evasion rate of 74.4\% and 97.7\% against GBDT and MalConv classifiers, respectively. Against the commercial AVs, the MAB agent's evasion rate dropped to 48.3\%.

More RL-bassed attacks were proposed by Kozak et al. \cite{kozak2022generation}. The authors trained a DQN agent with similar modifications as in \cite{song2022mab} against the GBDT and MalConv classifiers. The adversarial malware examples generated by the DQN achieved 68.64\% and 13.32\% evasion rates against GBDT and MalConv, respectively. While their results are significantly worse than the results by Song et al. \cite{song2022mab}, to the best of our knowledge, Kozak et al. were the first to propose the reverse scenario of generating adversarial benign examples. The authors demonstrated the danger of benign AEs in evasion attacks by increasing the FPR of GBDT and MalConv models by 3.45\% and 14.29\%, respectively.

The remaining evasion attacks are a mixture of different approaches that do not fit into gradient and RL-based groups.

Hu et al. \cite{hu2017generating} demonstrated the capabilities of generative adversarial networks (GANs) in generating malware AEs. The generator network operated by modifying feature vectors representing API calls captured from malicious files. The discriminator network represented a substitute malware detector and was trained to detect feature vectors modified by the generator. The authors reported an evasion rate of 98\% to 100\% when the generated vectors were transferred and evaluated against other ML-based malware classifiers. Unfortunately, the authors did not propose a method of converting the feature vectors back to executable binaries, thus limiting the real-world application of their work.

Further, Demetrio et al. \cite{demetrio2021functionality} experimented with evolutionary algorithms to create malware AEs. The evolutionary algorithm was dedicated to solving an optimization problem balancing maximum evasion rate with minimal perturbation size. Evolutionary techniques such as selection, cross-over, and mutation are applied to vectors to show how adversarial benign content inside malicious PE files should be injected. The optimized feature vectors are later applied on genuine malicious files, creating malware AEs. The authors reported an evasion of 12 out of 70 detectors hosted on VirusTotal\footnote{\url{www.virustotal.com}} website.

\subsection{Poisoning Attacks}
Biggio et al. \cite{biggio2014poisoning} presented one of the first poisoning attacks in the domain of malicious software. The authors demonstrated how an attacker could subvert the behavior clustering process of Malheur \cite{rieck2011automatic}, an open-source tool, by injecting carefully crafted poisoning samples. Biggio et al. experimented with bridge-based (adding points to bridge clusters), random, and F-measure minimizing attacks. The results showed that poisoning the training dataset by as little as 5\% samples from the bridge-based attack can lead to complete deterioration of clustering performance where Malheur merges all samples to a single cluster (originally 40 clusters before poisoning attack).

Further, Chen et al. \cite{chen2018automated} investigated the vulnerability of ML-based malware detection systems to poisoning attacks on the Android platform. The authors introduced three types of attacks (weak, strong, and sophisticated) and used a customized adversarial crafting algorithm to generate crafted camouflage samples that misled classifiers. The proposed defense system, KuafuDet, includes an offline training phase and an online detection phase, intertwined through a self-adaptive learning scheme that uses similarity-based filtering to identify and retrain on suspicious false negatives. The authors first show that SVM-based detectors are susceptible to poisoning attacks by demonstrating up to a 30\% decrease in accuracy. Later, Chen et al. prototyped their retraining mechanism on suspicious samples to increase accuracy by at least 15\%.

Next, Sasaki et al. \cite{sasaki2019embedding} explored using data poisoning attacks to embed backdoors in malware detection systems. The proposed methodology involves generating poisoning data that misclassifies specific types of malware as benign software while maintaining the detection accuracy for other malware (so-called targeted poisoning). The attack framework consists of three steps: selecting backdoor malware, generating poisoning data using an optimization problem, and injecting the poisoning data into the training set. The authors introduced a constraint term to ensure the poisoning data resembles benign data, making it harder to detect. Logistic regression was used as the target malware detector. The result showed that the proposed method effectively increases the false negative rate for backdoor malware (over 80\% at 15\% contamination rate for selected malware) without significantly affecting the detection rates for other malware or benign software.

The work of Sasaki et al. \cite{sasaki2019embedding} was followed by Narisada et al. in \cite{narisada2020stronger}. The authors introduced two new targeted poisoning attack algorithms designed to evade common data sanitization defenses, specifically the sphere defense. The proposed methods include a basic attack that generates poisoning points by minimizing validation loss while ensuring points remain within a feasible set and a streamlined attack that combines label-flip attacks with the validation loss minimization approach. As previously, the logistic regression was used as a target classifier, and sphere defense was applied to remove 15\% of the points from the training data. The results showed that both proposed algorithms successfully evade the sphere defense, with the streamlined attack achieving a 91\% attack success rate at 15\% contamination rate.

\section{Conclusion} \label{sec:conclusion}
In this work, we explored a new scenario of benign AEs and their effectiveness in evasion and poisoning attacks. We utilized several well-known generators of adversarial malware and modified them to create benign examples.

The experimentation provided insights into the effectiveness of benign and malware AEs. In evasion attacks, specific generators were highly effective at producing benign AEs that are misclassified as malware by the target classifier in 97\% of cases. Although these benign AEs do not directly threaten the system, they cause an increase in FPR. This vulnerability can erode consumer faith in AV solutions and displease software vendors due to the blocking of their legitimate applications.

In a more realistic scenario where benign AEs are used to poison a training dataset of malware detectors, we found no substantial difference in the effectiveness of contaminating the training dataset with benign or malicious AEs produced by a single generator. This result suggests that the specific generator of benign or malicious AEs has little effect on the overall contamination when a single source is used. However, as mentioned in \nameref{sec:experiments_discussion}, we operated with a limited dataset size in this scenario, and increasing the available training samples could provide more practical results.

Nevertheless, a measurable difference was recorded when combining AEs from different generators for the poisoning attack. Our results show that including benign AEs in the training dataset outperforms malware or mixed AEs in reducing detection rates at fixed levels of FPR. These findings reveal a new pathway for poisoning attacks, requiring security engineers to remain vigilant for both benign and malware AEs infiltrating training datasets. 


\mypar{Future Work.} We envision that more research on the efficacy of benign AEs will follow. We plan to study the use of benign AEs in targeted poisoning attacks more profoundly and investigate how to create dedicated generators of benign AEs to contaminate training datasets more effectively.





\section*{Acknowledgements}
This work was supported by the Grant Agency of the Czech Technical University in Prague, grant No. SGS23/211/OHK3/3T/18 funded by the MEYS of the Czech Republic.

\bibliographystyle{plain}
\bibliography{references.bib}

\end{document}